\documentclass[twocolumn,epjc3]{svjour3}  

\smartqed
\RequirePackage{graphicx}

\begin{document}

\title{Detailed study of the decay of $^{21}$Mg}

\author{
  E.A.M. Jensen\thanksref{e1,addr1}
  \and S.T. Nielsen\thanksref{addr1}
  \and A. Andreyev\thanksref{addr8}
  \and M.J.G. Borge\thanksref{addr2}
  \and J. Cederk\"{a}ll\thanksref{addr3}
  \and L.M. Fraile\thanksref{addr4}
  \and H.O.U. Fynbo\thanksref{addr1}
  \and L.J. Harkness-Brennan\thanksref{addr9}
  \and B. Jonson\thanksref{addr5}
  \and D.S. Judson\thanksref{addr9}
  \and O.S. Kirsebom\thanksref{addr1}
  \and R. Lic\u{a}\thanksref{addr6}
  \and M.V. Lund\thanksref{addr1}
  \and M. Madurga\thanksref{addr7}
  \and N. Marginean\thanksref{addr6}
  \and C. Mihai\thanksref{addr6}
  \and R.D. Page\thanksref{addr9}
  \and A. Perea\thanksref{addr2}
  \and K. Riisager\thanksref{addr1}
  \and O. Tengblad\thanksref{addr2}
}

\thankstext{e1}{e-mail: ej@phys.au.dk}

\institute{Institut for Fysik \& Astronomi, Aarhus Universitet, DK-8000 Aarhus C, Denmark \label{addr1}
  \and School of Physics, Engineering and Technology, University of York, York YO10 5DD, N Yorkshire, UK \label{addr8}
  \and Instituto de Estructura de la Materia, CSIC, E-28006 Madrid, Spain \label{addr2}
  \and Department of Nuclear Physics, Lund University, SE-22100 Lund, Sweden \label{addr3}
  \and Grupo de Fisica Nuclear, EMFTEL \& IPARCOS, Facultad de Ciencias Fisicas, Universidad Complutense de Madrid, 28040 Madrid, Spain \label{addr4}
  \and Department of Physics, University of Liverpool, Liverpool L69 7ZE, UK \label{addr9}
  \and Department of Physics, Chalmers University of Technology, SE-412 96 G\"{o}teborg, Sweden \label{addr5}
  \and ``Horia Hulubei'' National Institute of Physics and Nuclear Engineering, RO-077125 Magurele, Romania \label{addr6}
  \and Department of Physics and Astronomy, University of Tennessee, Knoxville, Tennessee 37996, USA \label{addr7}
}

\maketitle

\begin{abstract}
  Beta-delayed proton and gamma emission in the decay of $^{21}$Mg has been measured at ISOLDE, CERN with the ISOLDE Decay Station (IDS) set-up.
  The existing decay scheme is updated, in particular what concerns proton transitions to excited states in $^{20}$Ne.
  Signatures of interference in several parts of the spectrum are used to settle spin and parity assignments to highly excited states in $^{21}$Na.
  The previously reported $\beta$p$\alpha$ branch is confirmed.
  A half-life of 120.5(4) ms is extracted for $^{21}$Mg.
  The revised decay scheme is employed to test mirror symmetry in the decay and to extract the beta strength distribution of $^{21}$Mg that is compared with theory.
\end{abstract}

\section{Introduction}
\label{intro}
The mechanism of beta-delayed particle emission provides an attractive means of probing the nuclear structure of neutron- and proton-deficient nuclei.
For the case of the neutron-deficient nucleus $^{21}$Mg with spin and parity $5/2^+$, the selectivity of beta decay enables the precise study of a subset of the otherwise large density of excited and relatively broad \cite{Fire15} states of $^{21}$Na.
The preferential population of $3/2^+$, $5/2^+$ and $7/2^+$ levels in $^{21}$Na in the decay of $^{21}$Mg reveals details of the sd-shell nucleus $^{21}$Na which are not easily accessible by other experimental means.

\begin{figure*}
  \includegraphics[width=\textwidth]{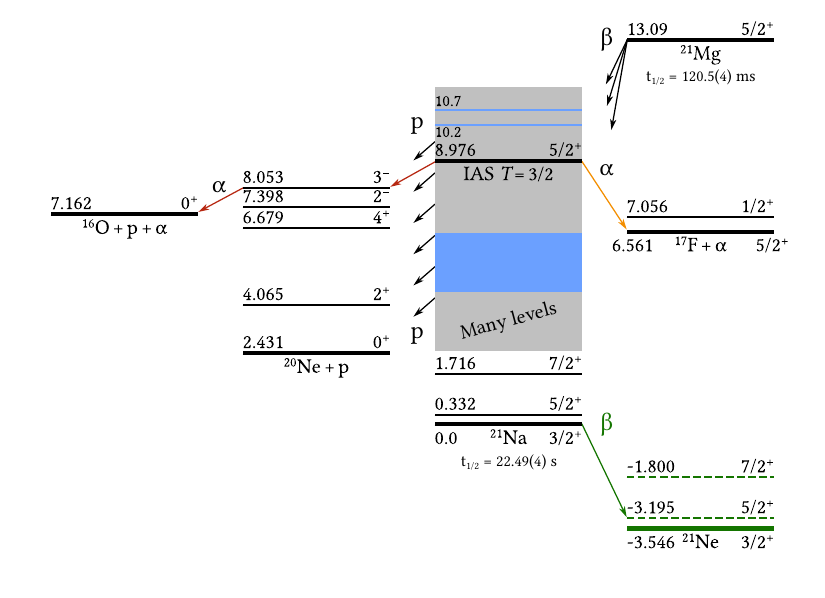}
  \caption{Decay scheme of $^{21}\mathrm{Mg}$.
  Energies are given in MeV relative to the ground state of $^{21}\mathrm{Na}$.
  The uncertainties of the energies are $\pm 1$ on the corresponding last significant figure.
  Not all individual levels in $^{21}\mathrm{Na}$ that take part in proton decay are shown explicitly; they are listed in table \ref{tab:peaks}.
  Regions of excitation energy in $^{21}\mathrm{Na}$ where $\beta$p transitions have not previously been identified are highlighted in light blue.
  $\beta$p$\alpha$ and $\beta \alpha$ transitions are highlighted with red and orange arrows respectively.
  $^{21}\mathrm{Na}$ also beta decays to $^{21}\mathrm{Ne}$, and gamma transitions between the indicated levels in $^{21}\mathrm{Ne}$ are observed in the data.}
  \label{fig:decay-scheme}
\end{figure*}

Following the beta decay of $^{21}$Mg, the emission of a proton and/or an alpha particle is possible, as is the deexcitation of excited states via gamma emission; see fig. \ref{fig:decay-scheme}.
The combined amount of detected particles from the decay of $^{21}$Mg has, over time, become increasingly complete.
In the first comprehensive study of the decay of $^{21}$Mg \cite{Sext73}, only beta-delayed protons were observed.
The observed spectrum was compared to shell model calculations.
Later, an unpublished experiment from GANIL \cite{Thom03} detected both protons and gamma rays in the decay, and branching ratios could be determined on an absolute scale.
Our earlier experiment \cite{Lun15} on the decay of $^{21}$Mg detected both beta-delayed protons and alpha particles.
Based on this experiment, several revisions and extensions of the decay scheme were carried out; in particular, a $\beta$p$\alpha$ branch was observed for the first time \cite{Lun15a}.
The identified proton lines were placed in a decay scheme partly based on energy relations and knowledge of the level scheme of $^{21}$Na, albeit without the coincident detection of $\gamma$ rays.
More recently, a measurement \cite{Wang18} with $^{21}$Mg ions implanted in a Si-detector recorded the proton spectrum as well as protons coincident with $\gamma$ rays from the first excited state in $^{20}$Ne.
Other unpublished experiments of similar type have also been performed \cite{Thom03,Ruot20} where more gamma rays have been recorded.
These experiments have given important clarifications to the decay scheme, but, due to the implanted source, the proton energy resolution is rather limited.

The aim here is to use our data to further improve the decay scheme by combining detection of $\gamma$ rays with charged particle detection in a number of Si-detector telescopes positioned in close geometry.
By employing our revised decay scheme, mirror symmetries in the decays of $^{21}$Mg and $^{21}$F are tested (following \cite{Riis23}), and the beta strength distribution of $^{21}$Mg is extracted.

\section{The experiment}
\label{sec:exp}

The experiment was carried out in 2015 with the aim of studying the decay of $^{20}\mathrm{Mg}$.
Those results have already been published in \cite{Lun16}.
As part of the calibration of the setup, data were also taken on the decay of $^{21}\mathrm{Mg}$; these data are analysed in the present paper.

\begin{figure}
  \includegraphics[width=\columnwidth]{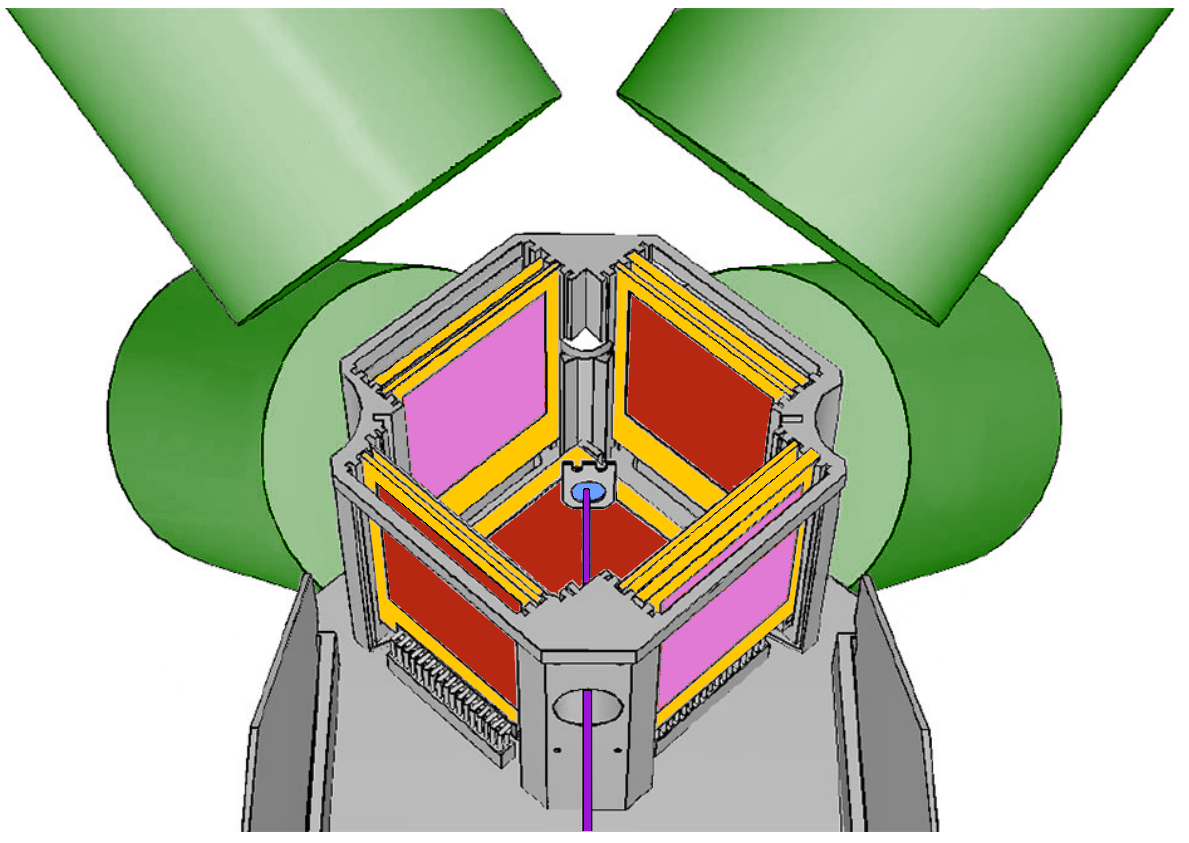}
  \caption{Schematic drawing of the setup used in the experiment.
  A 30 keV $^{21}\mathrm{Mg}$ beam goes through the front collimator and is stopped in a carbon foil.
  In the horizontal plane four $\Delta E$-$E$ silicon detector telescopes are placed and below the foil an additional silicon detector; the $\Delta E$ detectors are situated roughly 40 mm from the centre of the carbon foil, and the backing $E$ detectors are situated 5 mm further behind. The four cylinders outside the silicon detector holder represent the 4 High-Purity Germanium detectors which are four-fold segmented and situated roughly 30 cm from the carbon foil.
  This drawing is a slightly modified version of the one presented in \cite{Lun16}.}
  \label{fig:setup}
\end{figure}

A complete description of the beam production and experimental setup is given in \cite{Lun16}; here we give a brief summary.
A 30 keV $^{21}\mathrm{Mg}$ beam was produced at the ISOLDE facility at CERN \cite{Cat17} by bombarding a SiC target with a pulsed 1.4 GeV proton beam.
RILIS \cite{Val17} was employed to selectively laser-ionise $^{21}\mathrm{Mg}$ for extraction from the hot cavity ion source, connected to the SiC target, and HRS \cite{Cat17} was employed for subsequent mass-to-charge separation.
Due to surface ionisation of the target, a significant amount of the isobar $^{21}\mathrm{Na}$ was also present in the developed radioactive beam; an estimate of the resultant contamination is given at the end of this section.
From HRS, the radioactive beam was guided to the ISOLDE Decay Station (IDS) \cite{Fyn17}, where the low-energy radioactive beam was implanted in a carbon foil of thickness 24.5(5) $\mu\mathrm{g/cm^{2}}$.
The carbon foil was surrounded by silicon detectors, and, outside the vacuum chamber, by High-Purity Germanium (HPGe) detectors; a sketch of the setup is shown in fig. \ref{fig:setup}.
For the results presented in this paper, data from two of the four \mbox{$\Delta E$-$E$} silicon detector telescopes as well as the silicon detector below the foil are used (dark red detector surfaces in fig. \ref{fig:setup}).
Data from the remaining silicon detectors have been excluded due to technical issues during data taking of $^{21}\mathrm{Mg}$.
The utilised $\Delta E$ detectors are double-sided silicon strip detectors (DSSSDs) with 16$\times$16 strips spanning an area of 50$\times$50 $\mathrm{mm^{2}}$ with ultra-thin entrance windows \cite{Ten04}.
The utilised $E$ detectors are 50$\times$50 $\mathrm{mm^{2}}$ single-sided pad detectors.
Both $E$ detectors have active layer thicknesses of 500 $\mu$m and the $\Delta E$ detectors have active layer thicknesses of 42 and 67 $\mu$m.
The silicon detector below the foil is the same type of DSSSD as the two $\Delta E$ detectors with an active layer thickness of 1000 $\mu$m.
The 4 HPGEs situated outside of the chamber are four-fold segmented Clover detectors.

The setup used in this experiment allows for the identification of the protons, alphas and gammas emitted in the complex beta-delayed particle emission of $^{21}\mathrm{Mg}$; see fig. \ref{fig:decay-scheme}.
The segmentation  of the DSSSDs provides accurate information on a given particle's trajectory and, in turn, allows for accurate determination of initial particle energies, whilst the small thickness of the $\Delta E$ detectors suppresses beta response at low energy.
The segmentation of the Clover detectors provides improved detection efficiency via add-back correction of the gammas.
The inherent high resolution of the Clover detectors is not utilised fully in proton-gamma coincidences due to the data acquisition system of the experiment being optimised for resolution on the silicon detectors.
This is of less importance in our analysis since the relevant gamma lines are well separated as can be expected in a light nucleus.

The $\Delta E$\emph{-E technique} is well-established \cite{goulding,perry,Jen23}, and it is in principle straightforward to unambiguously identify various kinds of charged particles when they deposit characteristic fractions of their initial energies in both the $\Delta E$ and the $E$ detectors.
When the initial particle energies are less than the punchthrough threshold of a given $\Delta E$ detector, however, energy regions in which initial particle energies \emph{cannot} be uniquely assigned will emerge \cite{Jen23}.
In close geometry, i.e. when the angle of incidence with respect to the $\Delta E$ detector surface can vary significantly, these spurious energy regions can span many hundreds of keV.
Details of the manifestation of this effect and how it can be accomodated are given in \cite{Jen23}.
In the present work, energy loss corrections based on position information of the silicon detectors are employed, and the energy-dependent variation in detection efficiency, due to the manifestation of the mentioned spurious energy regions, is indicated in fig. \ref{fig:particle}.

Based on the revised $\Delta E$-$E$ analysis methods, and based on identification from previous measurements \cite{Lun15,Lun15a}, protons and alpha particles are reliably identified above and below the relevant punchthrough thresholds of our setup.

As the data of the present paper served as calibration data for the study of $^{20}$Mg, the silicon detectors were primarily energy-calibrated to protons emitted in the decay of $^{21}$Mg.
The reference proton energies employed for these calibrations were adopted from\ \cite{Fire15}.
Due to the low punch through threshold, with respect to protons, of the thinnest 42 $\mu$m silicon detector, this specific detector was, instead, energy-calibrated to alpha particles emitted in the decay of $^{20}$Mg.
The pulse height defect (see e.g.\ \cite{Kir14}) is accounted for by converting from alpha-calibrated energies $E_{\alpha}$ of the 42 $\mu$m silicon detector to proton energies $E_{\mathrm{p}}$ via the empirical relation $E_{\mathrm{p}} = 0.986(E_{\alpha} - 8\,\mathrm{keV})$.
The germanium detectors were energy- and efficiency-calibrated with a $^{152}$Eu source.
Further details on the calibrations of the silicon detectors can be found in\ \cite{Niel16}, and further details on the calibrations of the germanium detectors can be found in\ \cite{Lun16}.

About $n_{\mathrm{Mg}}=10^6$ decays of $^{21}\mathrm{Mg}$ were observed to proceed via charged particle emission during the measurement time of 5 hours.
Correcting for detection efficiencies of the charged particle detectors (fig. \ref{fig:particle}) and for our estimate of the $\beta$p branching ratio (sect. \ref{sec:disc}), roughly $N_{\mathrm{Mg}}=5 \times 10^7$ $^{21}\mathrm{Mg}$ ions were delivered to our chamber during the measurement time, and, on average, 3000 $^{21}\mathrm{Mg}$ ions were delivered to the chamber per second, corresponding to a production yield of 1600 ions per microcoulomb.

Based on the number of observed gamma transitions from the first excited state in $^{21}$Na and from the first excited state in $^{21}$Ne (fig. \ref{fig:decay-scheme}), we employ branching ratios for the gamma transitions and for the beta decay of $^{21}$Mg from literature, and we estimate that the ratio of decaying $^{21}$Na to decaying $^{21}$Mg in the carbon foil, $R=N_{\mathrm{Na}}/N_{\mathrm{Mg}}$, was around 5 for the data presented in this paper.
The branching ratios for the gamma transitions are adopted from\ \cite{Fire15}, and employing either the branching ratio for the beta decay of $^{21}$Mg of 67.4\% from\ \cite{Sext73} or of 77-78\% from the current work (sect. \ref{sec:disc}) does not significantly influence the resulting ratio $R$.

The gamma and charged particle data presented in this paper includes only data from within the first 1.2 seconds after production of the radioactive beam.
As $^{21}$Mg has a half-life of 120.5(4) milliseconds (sect. \ref{sec:half-life}), while $^{21}$Na has a half-life of 22.55(10) seconds \cite{Fire15}, this time gate enhances gamma signals from the decay of $^{21}$Mg above the ever-present gamma signals due to the decay of $^{21}$Na and due to background radiation in the ISOLDE Hall.
The deduced number of decaying $^{21}\mathrm{Mg}$ ions, $N_{\mathrm{Mg}}$, from the charged particle spectra and from the gamma spectra are internally consistent.

\section{Results}
\label{sec:2}
The results will be presented as follows. The determination of the half-life of $^{21}$Mg is described first.
Accounts are then given of the observed gamma lines, of the singles and gamma-coincident proton spectra and, finally, of the proton-$\alpha$ particle coincidences.
The derived decay scheme is then discussed in detail.

\subsection{Half-life}

The half-life of the precursor $^{21}\mathrm{Mg}$ is determined by identifying and counting protons emitted from excited states in the emitter $^{21}\mathrm{Na}$ at various times after production of the radioactive $^{21}\mathrm{Mg}$ beam at ISOLDE.
The proton events are gathered from two of the setup's detector telescopes and from one thick DSSSD.

\begin{figure}
  \includegraphics[width=\columnwidth]{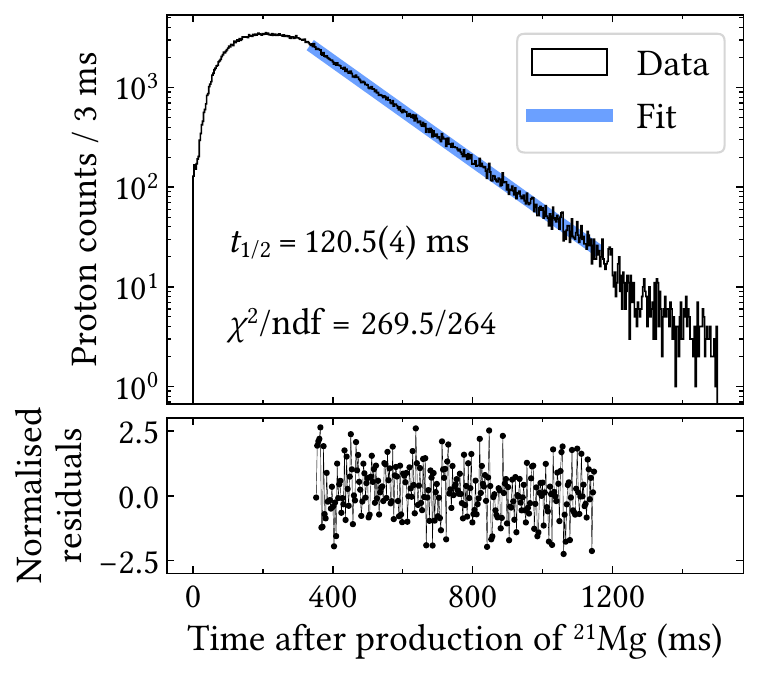}
  \caption{Time distribution of protons recorded from the decay of $^{21}$Mg.
  The shaded region marks the time interval used to deduce the halflife of $^{21}$Mg.}
  \label{fig:halflife}
\end{figure}

The number of proton counts detected at various times after production is shown in fig. \ref{fig:halflife}.
During the first 350 milliseconds the activity is led into the chamber and there is a build-up of (decaying) $^{21}\mathrm{Mg}$ within the chamber.
The subsequent exponential decay is characteristic of the half-life of $^{21}\mathrm{Mg}$.
The discontinuity at 1.2 seconds is due to the cycling of the Proton Synchrotron Booster at CERN, which resets the clock at integer multiples of 1.2 seconds.
The bump just before 1.2 seconds is due to signals from neutrons produced when the proton beam hits the production target, the reset of the clock only takes place after the separator high voltage is stable, see sect. 2.5 in \cite{Cat17}.
In the fit of the half-life we avoid this region and we do not include the region of lower statistics after 1.2 seconds.

The data in the region from 350 to 1150 milliseconds are fitted to a simple exponential decay using unbinned maximum likelihood.
For our Poisson-distributed data, the binned $\chi^2$ variable
\begin{equation}
  \chi^{2}_{\mathrm{P}} = 2\sum_{i}^{N} n_{i}\ln\frac{n_{i}}{\hat{m}_{i}}+\hat{m}_{i}-n_{i}
\end{equation}
follows, in the limit of a large sample size $N$, a $\chi^2$ distribution with $N - 2$ degrees of freedom (see e.g. \cite{Bak84}).
In the expression, $n_i$ is the observed number of counts in bin $i$ and $\hat{m}_{i}$ is the expected number of counts in bin $i$ estimated by the fit.
The result of the fit is a half-life $t_{1/2} = 120.5(4)$ milliseconds of $^{21}\mathrm{Mg}$, and, with the chosen binning shown in the figure, we extract a chi-square value of $\chi^2 = 269.5$ with 264 degrees of freedom.
Varying the end points of the fitting range with up to 50 milliseconds has no influence on the result at the given precision.
Our result is consistent with the current adopted value \cite{Kon21} of 120.0(4) milliseconds.

\subsection{The gamma spectrum}
\begin{figure*}
  \includegraphics[width=\textwidth]{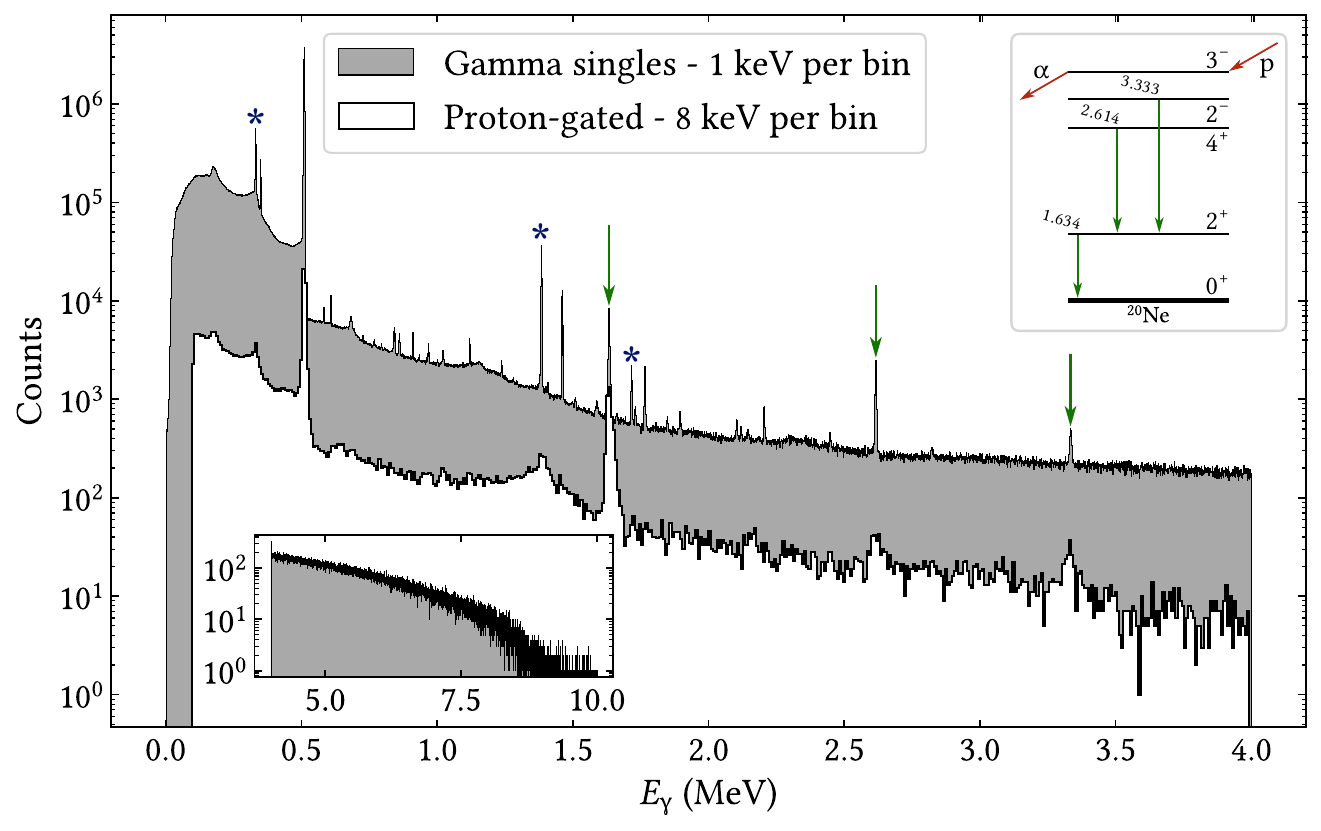}
  \caption{Combined gamma spectra of gamma energies $E_{\gamma}$ recorded in the four 4-fold segmented HPGe detectors, the high-energy part is shown in the lower inset.
  Both the singles and the proton-gated spectra are addback-corrected.
  The singles spectrum was recorded with a digital data acquisition system (DAQ) of high resolution, while the proton-gated spectrum was recorded with an analogue DAQ of lower resolution; hence the increased bin width and smaller resolution of the proton-gated spectrum.
  The upper inset shows the observed gamma transitions in $^{20}\mathrm{Ne}$ that are marked with arrows.
  Note that the 2.614 MeV line coincides in energy with a background line from $^{208}\mathrm{Tl}$.
  The gamma lines from bound state transitions in $^{21}\mathrm{Na}$ are marked by asterisks (*).
  See the text for further discussion.}
  \label{fig:gamma}
\end{figure*}

The combined gamma spectrum from all detectors including add-back is shown in fig.\ \ref{fig:gamma}.
As seen from fig.\ \ref{fig:decay-scheme} it may contain gamma rays from particle-bound states in $^{21}$Na and from states in $^{20}$Ne fed through proton-emission as well as from $^{21}$Ne (from the beta decay of $^{21}$Na).
The low-lying spectrum in $^{21}$Na is well-established and we observe, as has been done previously \cite{Thom03}, three gamma lines from the de-excitation of the two lowest-lying states in $^{21}$Na, below the proton separation energy: a 332 keV line from the first excited 5/2$^+$ level to the 3/2$^+$ ground state, a 1384 keV line from the second excited 7/2$^+$ level to the first excited 5/2$^+$ level, and a 1716 keV line from the 7/2$^+$ level to the 3/2$^+$ ground state; see fig.\ \ref{fig:decay-scheme}.
As was mentioned in section\ \ref{sec:exp}, there is evidence of direct production of $^{21}$Na from our target, as we observe quite a strong 351 keV line from the beta decay of $^{21}$Na which, otherwise, has a branching ratio of just 5\%\ \cite{Fire15}. Due to this direct production of $^{21}$Na, we are unable to extract the intensity of the ground state branch.
Several transitions in $^{20}$Ne are also observed and are marked with arrows in fig.\ \ref{fig:gamma}.

With a gate in the time distribution the transitions from the decay of $^{21}$Mg can be enhanced further above background.
The following relative intensities of gamma transitions are derived after efficiency correction \cite{Lun16}:
\begin{itemize}
  \item 1716 keV transition to 1384 keV transition, i.e.\ the deexcitations of the 7/2$^+$ level: the ratio is 0.067(3), consistent with but more accurate than the previous value \cite{Fire15} of 0.075(22)
  \item 1384 keV transition to 332 keV transition: the ratio is 0.30(2), indicating that the ratio of feeding to the 7/2$^+$ and 5/2$^+$ levels is 0.46(4) (the previous estimate of the ratio of 0.27(6) came from the mirror decay of $^{21}$F)
    \item 1634 keV transition to 1384 keV transition: the ratio is 0.42(2), it will be used below to relate the feedings to levels above and below the proton threshold
\end{itemize}

The transitions in $^{20}$Ne are enhanced in fig.\ \ref{fig:gamma} in the spectrum of gamma rays coincident with a proton recorded in the Si detectors.
The background level is clearly reduced and three transitions in $^{20}$Ne are observed in agreement with \cite{Ruot20}, the inset in the figure gives the relevant partial level diagram.
Earlier experiments \cite{Thom03,Wang18} have only observed proton spectra coincident with the 1634 keV line, our results will be shown in the next section.
Note that the majority of the feeding to the $4^+$ and $2^-$ states as well as some feeding to the $3^-$ state proceed through the $2^+$ state and therefore will give a 1634 keV de-excitation gamma ray.

In the proton-gated spectrum one clearly observes the recoil-broadening of the 1.634 MeV line due to the preceeding proton emission.
Similar broadening occurs for the other $^{20}$Ne lines, but the statistics are less for these cases, the branching ratios relative to that of the 1.634 MeV line are at most 0.10(3) and 0.07(2) for the 2.614 MeV and 3.333 MeV lines, respectively.
(A background line from $^{208}\mathrm{Tl}$ is also present at 2.614 MeV, so that only an upper limit can be given here.)
As explained below, the $3^-$ state is also populated by beta-delayed protons and gamma rays from its deexcitation are expected.
However, the $3^-$ state mainly decays via alpha particle emission and the most intense gamma line at 3.987 MeV has a branching of about 7\%.
As such, only a few coincidences would have been expected with the current data, taking into account the low detection efficiency at 3.987 MeV gamma energy, and we do not see any direct indications of deexcitation from the $3^-$ state via gamma emission.

\subsection{The particle spectrum}
\begin{figure*}
  \includegraphics[width=\textwidth]{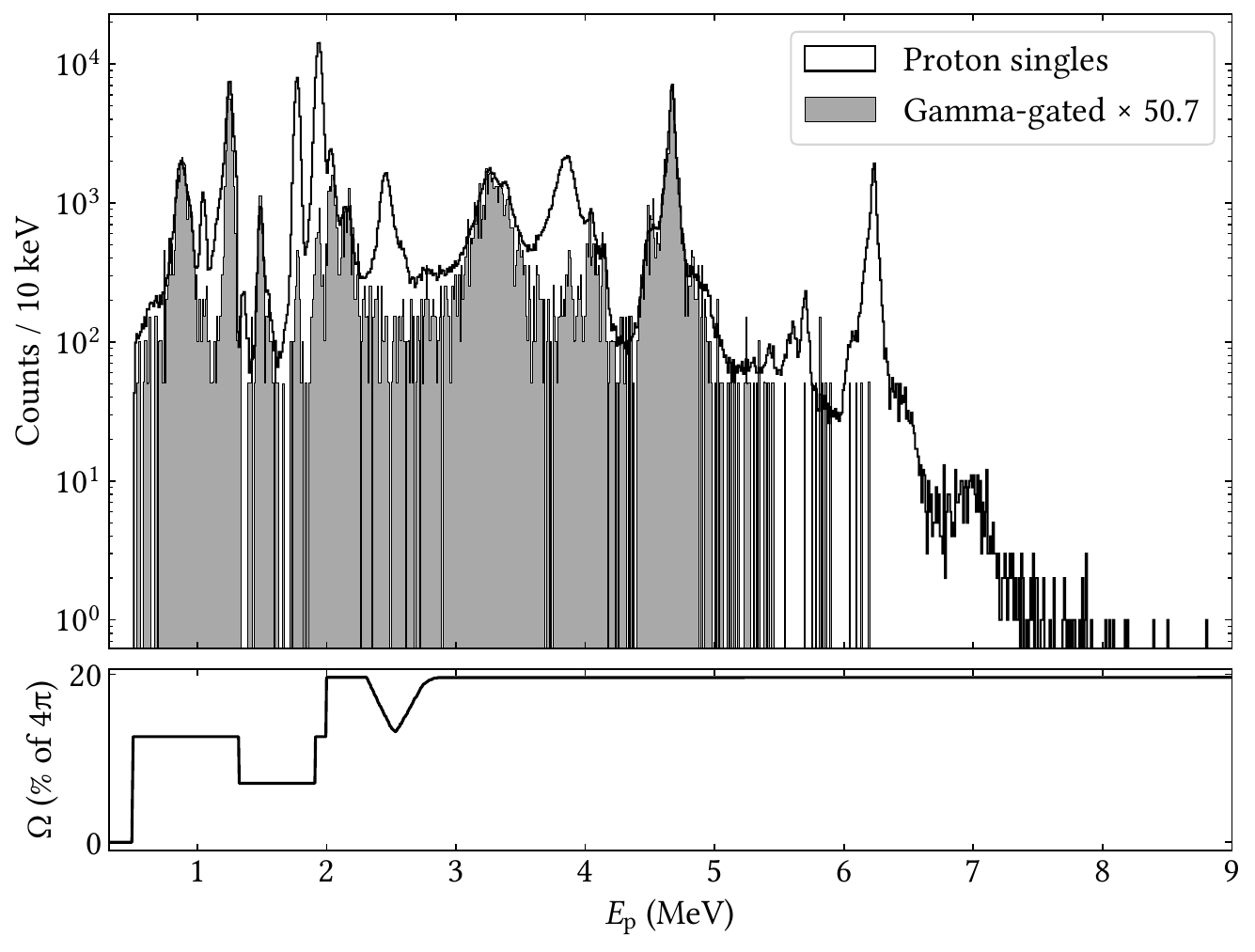}
  \caption{Combined charged particle spectrum recorded in the Si detectors is shown versus the proton energy (corrected for energy loss) $E_{\mathrm{p}}$.
  The shaded histogram shows which parts of the spectrum are in coincidence with the 1634 keV gamma transition; it is rescaled to match the singles spectrum.
  41(2) \% of the proton decays go to excited states in $^{20}$Ne; out of this roughly 2 \% and 4 \% of the protons go to the $4^+$ and $2^-$ states, and these states are deexcited via the $2^+$ state in $^{20}$Ne -- see the inset of fig. \ref{fig:gamma}.
  The lower panel shows the effective solid angle $\Omega$ of the setup at a given energy, as outlined in \protect\cite{Jen23}.}
  \label{fig:particle}
\end{figure*}

The events registered in the Si detector telescopes will mainly be $\beta$ particles at low energy and protons at higher energy.
There are also small contributions from the $\beta\alpha$ and $\beta$p$\alpha$ branches, however, protons will dominate the charged particle spectra where the $\beta\alpha$ and $\beta$p$\alpha$ branches contribute \cite{Lun15a}.
The strongest $\alpha$ line at 1954 keV is situated where the two most intense proton lines of the decay appear (fig.\ \ref{fig:particle}), and the other $\alpha$ lines of smaller energies are too weak to be disentangled from the proton spectrum without particle identification.
Overall, the contribution of the $\beta\alpha$ and $\beta$p$\alpha$ branches to the extracted $\beta$p spectra is negligible.

The final particle spectrum, obtained by combining the best spectra from the Si detectors, is shown in fig.\ \ref{fig:particle}.
We discard the two telescopes that have a 20 $\mu$m and 300 $\mu$m front strip detector (the first has too low resolution, the last appears to have not been fully depleted). The regions in the other detectors where punch-through protons or beta particles may contribute are also left out.
The bottom panel in fig.\ \ref{fig:particle} shows what the combined solid angle of the reliable detectors is at a given energy.
The sharp drop in the solid angle coverage around a proton energy $E_{\mathrm{p}}$ of 1.3 MeV, for example, results in a correspondingly sharp drop in the number of counts at the high-energy tail of the proton peak in the vicinity of this energy.
Details of the analysis procedure which lead to the illustrated variation in solid angle coverage are given in \cite{Jen23}.
The combined resolution only allows for determination of level widths above 50 keV.

The earlier experiments \cite{Thom03,Lun15,Wang18} have made clear that there is substantial proton feeding to excited states in $^{20}$Ne, as confirmed by our gamma spectrum shown in fig.\ \ref{fig:gamma}.
The parts of the proton spectrum that correspond to excited state transitions are identified with a gate on the 1634 keV $2^+ \rightarrow 0^+$ $\gamma$ transition, the spectrum obtained in this way is also displayed in fig.\ \ref{fig:particle} and is rescaled by the $\gamma$-ray detection efficiency of approximately 2 \%.
This spectrum is significantly improved in resolution compared to earlier experiments.

\begin{figure*}
  \includegraphics[width=\textwidth]{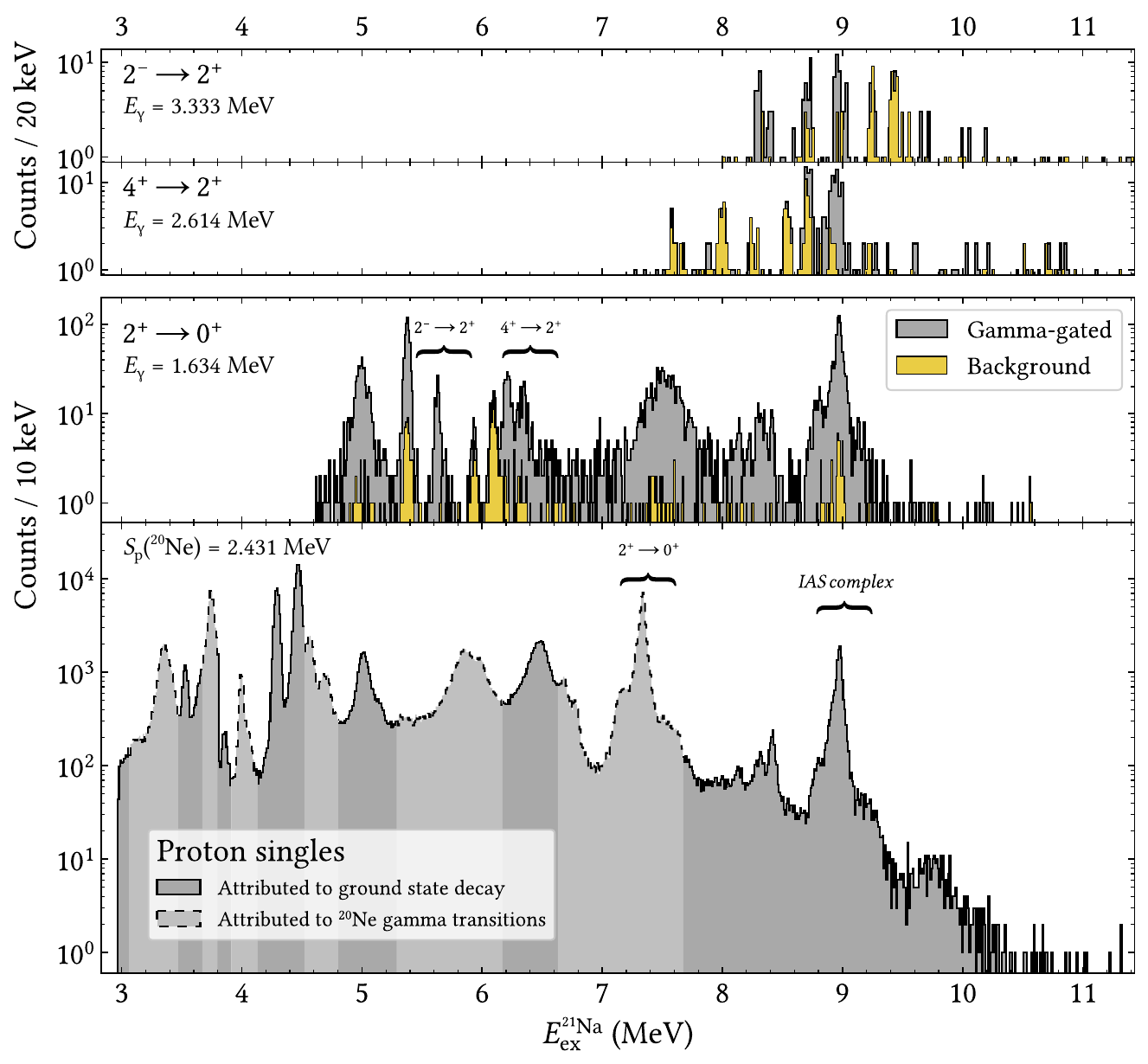}
  \caption{The beta-delayed proton spectra from the decay of $^{21}$Mg observed in singles (lowest panel) and with gates on the $\gamma$ rays at 1634 keV, 2614 keV and 3333 keV (upper panels) are shown displaced in energy so that all features appear at the appropriate excitation energy in $^{21}$Na, $E_{\mathrm{ex}}^{^{21}\mathrm{Na}}$.
  (Note that the $2^+$ gated spectrum will also contain lines from the $4^+$ and $2^-$ spectra as the gamma decays of these levels proceed through the $2^+$ level.)
  The background histograms in the upper panels are spectra obtained with a shifted gate in $\gamma$-ray energy and correspond to random coincidences.
  The curly brackets mark regions where decays from the IAS (and states close to it) occur.
  Note that the background lines do not resemble actual populations of levels at the indicated excitation energies:
  For example, the two background lines just above and just below the IAS in the upper two panels are actually due to the two most intense
  ground state transitions seen in the bottom singles spectrum at 4.3 and 4.5 MeV excitation energy.}
  \label{fig:comb_p}
\end{figure*}

The proton spectra resulting from gating on all three observed lines, as well as background spectra found with a displaced gate near the relevant gamma lines, are shown as a function of the deduced excitation energy in $^{21}$Na in fig.\ \ref{fig:comb_p}.
A more detailed account of this coincidence analysis can be found in \cite{Niel16}. The only clear feeding of the $4^+$ state is from the IAS in $^{21}$Na, but there seems to be an excess of coincidences from the unresolved region above the IAS.
In contrast, the $2^-$ state is fed both from the IAS and a state around 8.3 MeV, and the $2^+$ state is fed by quite a few states in $^{21}$Na.
The random coincidences seen with the displaced gates mainly occur for the most intense peaks in the singles proton spectrum.

Due to the low statistics in the coincidence data and the increase in uncertainty of our gamma ray efficiency at high gamma energy, we use the coincidence spectra mainly to place transitions correctly in the decay scheme and evaluate the intensity of the peaks from the singles spectrum.

\subsection{The proton-$\alpha$ particle coincidences}
\begin{figure}
  \includegraphics[width=0.45\textwidth]{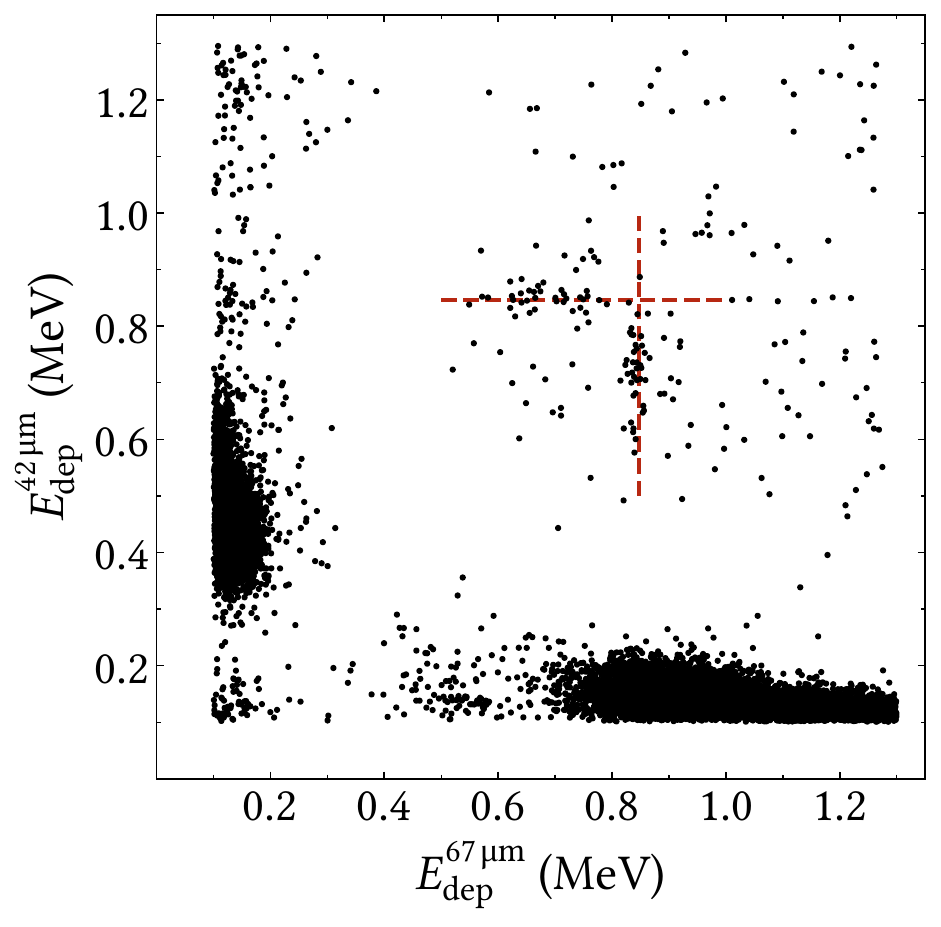}
  \caption{Coincidence spectrum of deposited energies $E_{\mathrm{dep}}$ between the two thin (42 $\mu$m and 67 $\mu$m) Si strip detectors.
  Apart from beta-proton coincidences along the energy axes and scattered random coincidences, the main feature is the events along the dashed lines that are interpreted as proton-alpha coincidences.
  See the text.}
  \label{fig:p_alpha}
\end{figure}

Our earlier experiment \cite{Lun15a} gave evidence for a $\beta$p$\alpha$ branch from $^{21}$Mg that was interpreted as proceeding through the $3^-$ state in $^{20}$Ne.
The energies of the $\alpha$ events pointed to this interpretation, but only a few coincidence events were recorded.
In order to verify this observation and interpretation, we show in fig.\ \ref{fig:p_alpha} the low-energy coincidence events recorded between the two opposing double-sided Si strip detectors with thicknesses of 42 $\mu$m and 67 $\mu$m.
The $\beta$ particles deposit only a few tens of keV in these detectors so we mainly see coincidences between heavy charged particles.
The only true coincidence events expected are from the $\beta$p$\alpha$ decays, but random coincidences with the strongest proton lines will also occur.
We observe, in the 2D plot of fig.\ \ref{fig:p_alpha}, about 150 low-energy events, close to the dashed lines in the figure.
Both detectors give the same projected energies (not corrected for energy losses in the collection foil and detector deadlayers), namely a sharp line around 850 keV and a broad distribution from a bit above 500 keV to just below 800 keV. This fits perfectely with the interpretation: A proton emitted from the IAS in $^{21}$Mg to the $3^-$ state in $^{20}$Ne will have a laboratory energy of 877 keV, the $\alpha$ particle emitted to the $^{16}$O ground state a laboratory energy of 714 keV, and the maximum recoil shift of the $\alpha$ energy will be 158 keV.
This is all consistent with our data.
One can in principle determine in which order two different particles are emitted from the amount of recoil broadening, but the final difference for different orderings is small and the smearing from energy losses will here be more important.

We conclude that the $\beta$p$\alpha$ observation in \cite{Lun15a} is confirmed.

\subsection{Derived decay scheme}
The interpretation of the proton spectrum has evolved since the first published data \cite{Sext73}, in particular what concerns broad features in the spectrum.
Before describing our findings, it may be useful to outline which complementary information can be used.

Allowed beta decay will populate levels in $^{21}$Na with spin-parity $3/2^+$, $5/2^+$ and $7/2^+$.
It is likely that the decay in a first approximation can be described within the sd-shell and we may therefore compare to the shell-model predictions given in \cite{Brow85}.
In the region up to the position of the IAS they predict 7, 7 and 6 levels for the three possible spin values (excluding the IAS itself).
We shall compare to the theoretical strength distribution later.

Several reaction experiments have provided a quite extensive knowledge of the level structure of $^{21}$Na, mainly from the reactions $^{20}$Ne(p,p), $^{20}$Ne(p,$\gamma$) and $^{20}$Ne (d,n), but also from reactions such as $^{23}$Na(p,t).
The latest compilation can be found in \cite{Fire15} where also information on the mirror nucleus $^{21}$Ne is available.
In the region up to the IAS (excluding levels mainly seen in earlier $^{21}$Mg decay experiments) the number of identified $3/2^+$, $5/2^+$ and $7/2^+$ levels in $^{21}$Ne and $^{21}$Na are 9, 6 and 3 (with 3--4 more tentative $7/2^+$ levels), and 8, 6 and 1, respectively.
We can therefore to a large extent base our interpretation on previous experiments for $3/2^+$ and $5/2^+$ levels, but not for $7/2^+$ levels which are difficult to populate in a one-step reaction from the $0^+$ ground state of $^{20}$Ne.

\subsubsection{The proton spectrum}
In fig.\ \ref{fig:particle}, the singles spectrum shows all of the observed proton groups from the decay, and the gamma gated spectrum reveals which proton groups are due to decays to excited states in $^{20}$Ne.
This information is compiled into fig.\ \ref{fig:comb_p}, in which the population of the various excited states in $^{21}$Na is depicted, and the various proton groups are attributed to decays either to the ground state or to excited states in $^{20}$Ne.
We first go through the energy spectrum and note differences to earlier work as well as regions where contributions from excited state transitions reduce the sensitivity to ground state transitions.

Of the four peaks in the region up to 1.7 MeV (laboratory proton energy), only the peak at 1 MeV goes to the ground state.
The 1.5 MeV peak goes to the $2^-$ state in $^{20}$Ne and the other two to the $2^+$ state, except that the (slightly asymmetric) 0.9 MeV peak has contributions also from transitions to the $2^-$ and $3^-$ states.
Our solid angle coverage changes in the region from 1.3 MeV to 1.7 MeV, see fig.\ \ref{fig:particle}, and we do not completely resolve the small extra peaks in this region observed in \cite{Lun15,Wang18}.

The two strongest peaks, at 1.8-2 MeV go to the ground state, while the two following peaks go to the $2^+$  and $4^+$ states.
The broad slightly asymmetric peak at 2.5 MeV goes again to the ground state, there is unresolved ground state strength up to 3 MeV followed by a quite broad complex between 3 MeV and 3.5 MeV that earlier \cite{Sext73} has been fitted with up to four peaks, but now (as in \cite{Lun15}) is interpreted as a broad peak leading to the $2^+$ state and a smaller narrow peak on the high energy side that presumably goes to the ground state.

The region 3.5--4.3 MeV again features a broad peak that goes to the ground state, with two peaks at higher energy (and one slightly lower only seen in the coincidence spectrum) that go to the $2^+$ state, again more consistent with \cite{Lun15} than with \cite{Sext73,Wang18}.
The strength in the 4.3--5.0 MeV region goes mainly to the $2^+$ state and shows structure both below and above the main peak, as in \cite{Lun15,Wang18}.
The 5.2--6 MeV region contains in the middle three peaks going to the ground state with indications below and above for strength to the $2^+$ state; this is a region where the interpretation has differed earlier \cite{Sext73,Wang18,Fern81}.
The spectrum above 6 MeV contains decays to the ground state.

In the regions 3--3.5 MeV and 4.3--5 MeV we are not sensitive to weak peaks going to the ground state.

\subsubsection{The excitation energy spectrum}
The next step is to systematically go through the experimental excitation energy spectrum in  fig.\ \ref{fig:comb_p}.
When transitions to several states in $^{20}$Ne are observed (typically the ground state and the $2^+$ state) the relative decay rates can be extracted.
Finally, the observed peaks can be compared to the existing level schemes \cite{Fire15} and spin values may be assigned.
Some assignments are necessarily tentative, for example for the energy region above the IAS where the statistics is low.

\begin{figure}
  \includegraphics[width=0.48\textwidth]{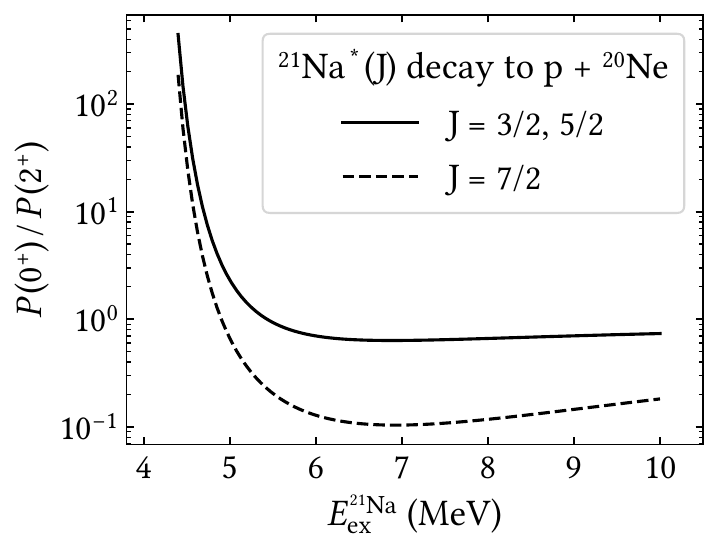}
  \caption{
  The ratio of penetrabilities for transitions into the $0^+$ ground state, $P( 0^{+} )$, and $2^+$ first excited state, $P( 2^{+} )$, in $^{20}$Ne as a function of excitation energy in $^{21}$Na, $E_{\mathrm{ex}}^{^{21}\mathrm{Na}}$, and for different spin values, $J$.
  $P( 0^{+} ) / P( 2^{+} )$ is plotted against $E_{\mathrm{ex}}^{^{21}\mathrm{Na}}$ in order to take the smaller proton kinetic energies to the $2^+$ state in $^{20}$Ne, as compared to the $0^+$ state, into account.
  }
  \label{fig:penet}
\end{figure}

The proton penetrabilities enter in the relative decay rates to states in $^{20}$Ne.
For $3/2^+$ and $5/2^+$ levels in $^{21}$Na the protons will be in d-waves to the ground state and $4^+$ state in $^{20}$Ne and in s-waves to the $2^+$ state.
For $7/2^+$ levels the protons are in g-waves to the ground state, d-waves to the $2^+$ and s-waves to the $4^+$ state.
For both negative parity states and all levels p-wave emission suffices.
Figure \ref{fig:penet} shows the ratio of penetrabilities to the two lowest states, the $2^+$ feeding is seen to start around 5 MeV and is then quickly favoured for all spin values.

The proton lines observed in the current experiment are collected in table \ref{tab:peaks} as a function of deduced level energy.
The estimated width and branching ratios to the ground state and $2^+$ state in $^{20}$Ne are listed and preferred spins are given.
In several cases, including the levels at 3544 keV, 4294 keV, 4468 keV and the IAS at 8976 keV, the identification to assigned levels in the literature \cite{Fire15} (also given in the table) is unambiguous and the spin can be safely taken over.
Comments are given below on other cases.

\begin{table*}
  \caption{Beta-delayed protons from $^{21}$Mg listed by their corresponding intermediate level excitation energies in $^{21}\mathrm{Na}$, $E_{\mathrm{ex}}^{^{21}\mathrm{Na}}$, and their widths $\Gamma$.
  The corresponding proton energies for transitions to the $0^+$ ground state of $^{20}$Ne, $E_{\mathrm{p}}^{0^+}$, are also given in the far left column.
  Results from the current experiment are listed in the left block along with deduced branching ratios $b_\mathrm{p}$ to the $0^+$ ground state and $2^+$ first excited state of $^{20}\mathrm{Ne}$ as well as spin $J$ assignments of the intermediate levels in $^{21}\mathrm{Na}$.
  Literature values from \protect\cite{Fire15} are listed in the middle block along with their spin and parity $J^\pi$ assignments.
  Branching ratios to the $0^+$ ground state of $^{20}\mathrm{Ne}$, $b_0$, from \protect\cite{Fern81} are listed in the right block.}
  \label{tab:peaks}
  \begin{tabular}{llllll|lll|l}
    \hline\noalign{\smallskip}
    \multicolumn{6}{c|}{Current experiment} & \multicolumn{3}{c|}{Literature, \protect\cite{Fire15}} & Literature, \protect\cite{Fern81} \\
    $E_{\mathrm{p}}^{0^+}$ & $E_{\mathrm{ex}}^{^{21}\mathrm{Na}}$ & $\Gamma$$^a$ & \multicolumn{2}{c}{b$_{\mathrm{p}}$ (\%)$^b$} & $J$$^c$ & $E_{\mathrm{ex}}^{^{21}\mathrm{Na}}$ & $\Gamma$ & $J^{\pi}$ & b$_0$ (\%) \\
    (MeV) & (MeV) & (keV) & $0^+$ & $2^+$ & & (keV) & (keV) & & \\
    \noalign{\smallskip}\hline\noalign{\smallskip}
    1.04(2)   & 3.52(2)     & S & 1.5(2) & n.a. & L & 3544.3(4) & 0.0155(14) & $5/2^+$ & \\
    1.36(1)   & 3.86(1)$^d$ & S & 0.66(3) & n.a. & L & 3862.2(5) & 0.0026(3) & $5/2^-$ & \\
    1.77(2)   & 4.29(2)     & S & 16.5(3) & $-$ & L & 4294.3(6) & 0.0039(1) & $5/2^+$ & \\
    1.93(2)   & 4.46(2)     & S & 23.2(5) & $-$ & L & 4467.9(7) & 0.021(3) & $3/2^+$ & \\
    2.47(1)   & 5.02(1)     & 110(15) & 4.6(3) & 4.4(4) & 3/2, 5/2 & & & & \\
    (2.59(2)) & (5.15(2))   & S & 0.3(2) & 0.3(2) & (3/2,5/2) & & & & \\
    2.80(1)   & 5.37(1)     & S & $<0.4$ & 10.8(4) & (7/2) & & & & \\
    3.35(2)   & 5.95(2)     & S & 0.15(5) &  & 3/2 & & & & \\
    3.59(1)   & 6.20(1)     & S &  & 1.5(4) & (7/2) & & & & \\
    3.85(2)   & 6.47(2)     & 130(25) & 6.5(6) & 0.5(3) & L & 6468(20) & 145(15) & $3/2^+$ & 90(12) \\
    4.82(2)   & 7.49(2)     & 200(50) & $<1.4$ & 7.2(6) & (L)$^d$ & 7609(15) & 112(20) & $3/2^+$ & 11(3) \\
    5.43(2)   & 8.13(2)     & S & 0.16(3) & 0.5(3) & L & 8135(15) & 32(9) & $5/2^+$ & 18(5) \\
    5.60(2)   & 8.31(2)     & S & 0.20(2) & 0.6(2) & L & 8397(15) & 30(13) & $3/2^+$ & 11(6) \\
    5.71(2)   & 8.42(2)     & S & 0.23(2) & 0.4(2) & L & 8464(15) & 25(9) & $3/2^+$ & 13(3) \\
    5.83(2)   & 8.55        & $^d$ &  &  & & & & & \\
    6.07      & 8.8         & $^d$ &  &  & & 8827(15) & 138(16) & $5/2^+$ & 28(5) \\
    6.23(1)   & 8.97(1)     & S & 2.10(3) & 8.4(3) & L & 8976(2) & 0.65(5) & $5/2^+$ & \\
    6.26      & 9.0         & $^d$ &  &  & & 8981(15) & 23(16) & $5/2^+$ & 8(4) \\
    7.0(1)    & 9.8(1)      & 300(100) & 0.06(1) & 0.15(2) & (L) & 9725(25) & 256(29) & $3/2^+$ & 53(7) \\
    7.4       & 10.2        & S & 0.008(3) & 0.07(2) & & & & & \\
    7.88      & 10.70(2)    & S & 0.002(1) &  & (3/2)$^d$ & & & & \\
    \noalign{\smallskip}\hline
  \end{tabular}

  $^a$S denotes a width up to our combined resolution of 50 keV.

  $^b$The fraction of the total beta-delayed proton spectrum.

  $^c$L denotes that the spin is taken from the literature. In allowed beta decay the level will have positive parity.

  $^d$See the discussion in the text.
\end{table*}

Recently, in a $^{24}$Mg(p,$\alpha$) experiment \cite{Kim21}, the broad level at 5 MeV has been observed at 5.036(11) MeV with a spin assignment of (3/2,5/2)$^+$.
It had about equal intensity for the $2^+$ and ground state branches as also observed here. The peak is slightly asymmetric on the high-energy side which could indicate the presence of a weakly fed level, it is listed in parentheses in the table.

The level at 5.37 MeV is mainly seen to proceed to the $2^+$ state, with at most a small ground state branch.
This and the fact that it has not been observed in proton scattering on $^{20}$Ne is compatible with a spin value of 7/2.
Note that the level at 5.6 MeV in the $2^+$-gated proton spectrum is from the IAS decay to the $2^-$ state and appears here due to the gamma cascade in $^{20}$Ne.
Futhermore, there is an indication for unresolved strength to the ground state from the 5 MeV level up to around 5.6 MeV excitation energy.
There is also a weak peak at 5.95 MeV that appears to proceed only to the ground state.

In a similar way the peak in the $2^+$-gated proton spectrum at 6.4 MeV arise from IAS decays to the $4^+$ state.
The level at 6.20 MeV mainly decays to the excited state and cannot have spin 3/2 since it does not interfere with the broad 6.47 MeV state, as for the 5.37 MeV level a spin value of 7/2 is likely.
The 6468 keV state is well-established and has spin-parity $3/2^+$.
A $^{20}$Ne(p,p) experiment with polarized protons \cite{Fern81} that measured elastic to total widths showed that the level decays mainly to the ground state, in agreement with our results; the proton branching ratio of levels to the $^{20}$Ne ground state are also displayed in the table.

The broad level seen clearly at 7.5 MeV in the $2^+$-gated proton spectrum may also have a ground state transition that in the singles spectrum lies below the IAS complex decaying to the $2^+$ state.
However, only the upper tail between 7.6 MeV and 7.8 MeV seems visible, so a proper extraction is not feasible.
The properties of the level seem close to the ones of the 7609 keV level in \cite{Fern81}, but the assignment is not firm.
The $^{20}$Ne(p,p) experiment  \cite{Fern81} reported 8 levels with spin-parity $3/2^+$ or $5/2^+$ in the region 8--9 MeV.
We tentatively see the first three of these levels and observe essentially featureless strength around 8.6 MeV that cannot be attributed easily to any (combination of) level(s), but clearly see much strength in connection with the IAS.
The interpretation of the decay of the IAS and the closely surrounding levels will be done in the following subsection.

The spectral shape above 9.2 MeV differs from the one reported in \cite{Lun15,Wang18}.
We see a clear indication for a peak at 9.8 MeV that tentatively is identified as the 9725 keV $3/2^+$ level seen earlier.
The only partially resolved strength between 10 MeV and 11 MeV that is present in both ground state and excited state spectra is here attributed to two levels.
The lower one is assumed narrow due to the signature in the $2^+$-gated spectrum and would then preferentially decay into the $2^+$ channel.
The upper one could be the, so far missing, $3/2^+$ level of isospin $T=3/2$ known \cite{Fire15} to exist in the other members of the multiplet, but this assignment is tentative and is mainly based on its small apparent width and its position about 1.7 MeV above the lowest $T=3/2$ level.

\subsubsection{The IAS decay and interference effects}
The $5/2^+$ IAS  at 8976 keV is the lowest isospin 3/2 state and is known \cite{Wilk92} to be narrow, $\Gamma =$ 650(50) eV with a proton width to the $^{20}$Ne ground state of $\Gamma_p =$ 117(10) eV corresponding to a partial branching ratio of $b_0 =$ 18(2)\%.
By combining our data with the results in \cite{Lun15,Lun15a}, we have the most extensive overview of the decay channels of the IAS.
Our results are shown in Table \ref{tab:IAS}, but we first discuss a few observations.

It is striking from figure \ref{fig:comb_p} that the levels close to the IAS appear to decay in about the same proportion to the $^{20}$Ne ground state and first excited state.
Even though the statistics is limited, the decays to the $4^+$ and $2^-$ states could also proceed in the same way.
We further reinterpret the $\alpha$-particle spectrum shown in \cite{Lun15a} as consisting of decays only to the $^{17}$F ground state via s-wave emission, and then see the same pattern appearing also for $\alpha$ decays.
(Our upper limit on the intensity of the 495 keV gamma ray in $^{17}$F is 1.4\% of the total proton intensity and is not sufficiently strong to distinguish the two interpretations.)
Assuming this is a correct interpretation, and since we cannot experimentally separate the IAS decays from the ones of the closely lying neighbouring levels, we will look at the combined decay of these levels, called ``the IAS complex''.

The relative branching intensities in table \ref{tab:IAS} are taken from table \ref{tab:peaks} except for the $\alpha$ and the $3^-$ branch that is rescaled from the results in \cite{Lun15,Lun15a}, the latter further being corrected for the known 93(3)\% $\alpha$ decay probability of the $3^-$ state.
Our extracted $b_0$ of 14(3)\% is slightly lower than the one from \cite{Wilk92} which could be due to an underestimation of the background under the IAS complex to the $2^+$ state.
We follow \cite{Sext73} and quote also the penetrabilities for the different channels, but do not see any obvious pattern in the results.

\begin{table}
  \caption{Decay channels of the IAS complex.
  Protons decay to states in $^{20}\mathrm{Ne}$; alpha particles decay to states in $^{17}\mathrm{F}$.
  Relative intensities $I$ and penetrabilities $P$ are given.
  Relative intensities are from table \ref{tab:peaks} and \cite{Lun15,Lun15a} (see text).}
  \label{tab:IAS}
  \begin{tabular}{lllll}
    \hline\noalign{\smallskip}
    Particle & Energy (MeV) & Final state & $I$ (\%) & $P$ \\
    \noalign{\smallskip}\hline\noalign{\smallskip}
    p       & 6.23(1) & $0^+$ (g.s.) & 2.10(3) & 1.22 \\
            & 4.67(1) & $2^+$ & 8.4(3) & 1.74 \\
            & 2.18(1) & $4^+$ & 1.4(3) & 0.088 \\
            & 1.50(1) & $2^-$ & 2.4(2) & 0.13 \\
            & 0.87(1) & $3^-$ & 0.07(1) & 0.015 \\
    $\alpha$ & 1.95(1) & $5/2^+$ (g.s.) & 0.23(2) & 0.11 \\
  \noalign{\smallskip}\hline
  \end{tabular}
\end{table}

There are at least two more $5/2^+$ levels that according to literature \cite{Fire15}, see table \ref{tab:peaks}, lie closer than one halfwidth to the IAS, so one must expect pronounced interference effects.
We note further the remark in \cite{Wilk92} that it was not possible to make a unique phase-shift analysis of the $T=1/2$ states in the region and therefore may not be able to trust the previous positions of the other $5/2^+$ levels.
However, it is clear that several $5/2^+$ levels appear in close vicinity to the IAS.
A complete elucidation would require fitting with the multi-level, multi-channel R-matrix theory as outlined e.g.\ in \cite{Bark88}.
Our statistics for the excited state channels do not allow a detailed investigation to be carried out, but we can illustrate the effects in a simple one-channel model.
(Technically, the energy dependence of the penetrability and shift factor will be neglected.)

\begin{figure}
  \includegraphics[width=0.48\textwidth]{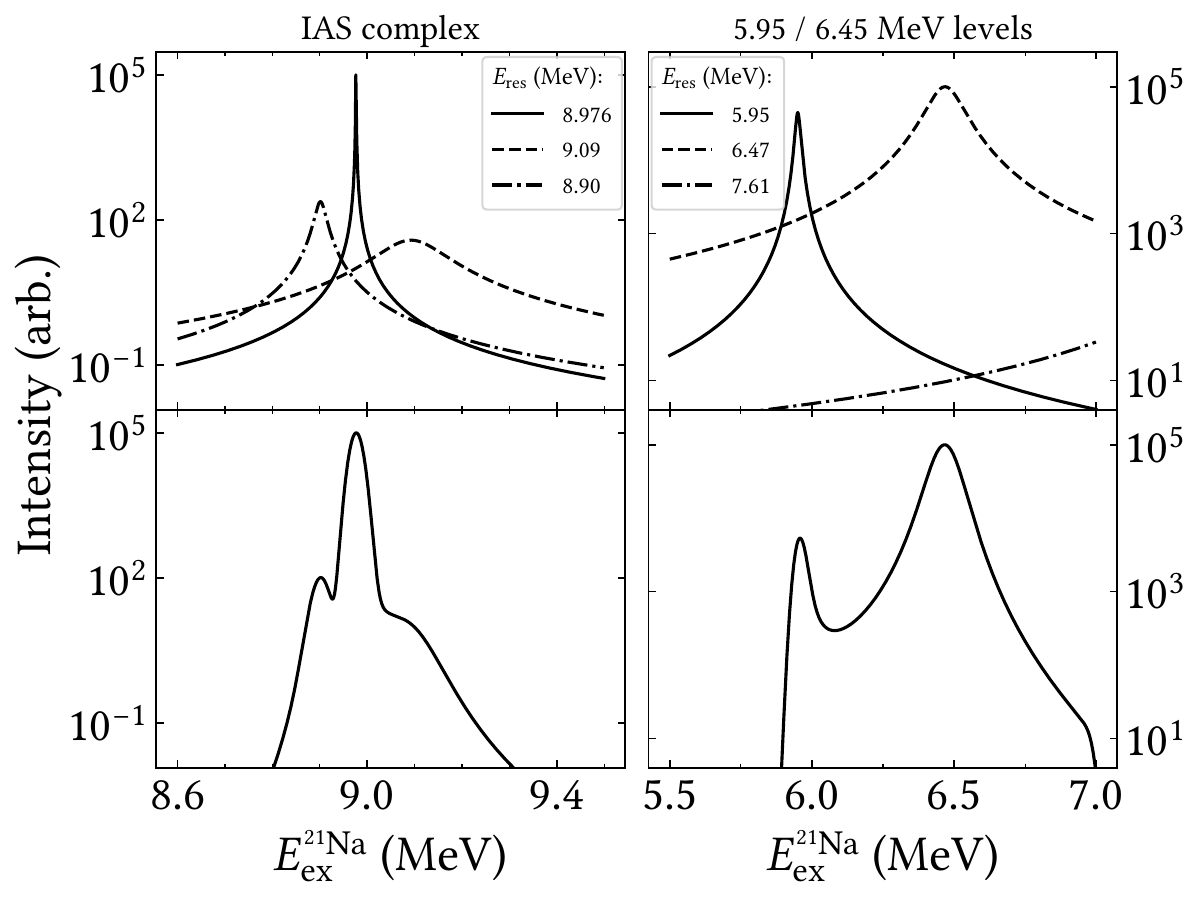}
  \caption{Sketches of the proton spectral shapes against excitation energy in $^{21}\mathrm{Na}$, $E_{\mathrm{ex}}^{^{21}\mathrm{Na}}$, of the IAS complex (left panels) and the 5.95 MeV and 6.45 MeV levels (right panels) for the non-interferring resonances of resonance energies $E_{\mathrm{res}}$ (top panels) and the interferring (bottom panels) cases.}
  \label{fig:interference}
\end{figure}

Figure \ref{fig:interference} shows in the left hand side the results for the IAS complex for the case without and with interference, respectively, where the literature widths (see table \ref{tab:peaks}) are used.
The IAS is so narrow that any interference effects will be very close to its position, and the interference dip is here placed at its lower side, in accordance with the pattern observed in \cite{Lun15}.
The other two levels have to interfere constructively between their positions and destructively outside in order that the outer edges of the IAS complex become as sharply defined as seen experimentally.
To reproduce the observation that the lower edge is sharpest, we have put the widest level above the IAS, opposite to the literature.
In the simulations, the IAS is at 8.976 MeV, the 23 keV wide level at 8.90 MeV and the 138 keV wide level at 9.09 MeV.
The strengths are adjusted to roughly correspond to the ground state IAS complex.

All three levels of the IAS complex can in principle be populated by Gamow-Teller transitions\footnote{A theoretical estimate of the Gamow-Teller strength to the IAS is 0.33; see sect. \ref{sec:disc}.},
and a small part of the Fermi feeding may also go to the two less intense levels as isospin-mixing is known to occur
from the small proton emission width of the IAS.
In both cases interference between the three levels will take place.
A better energy resolution would allow for a more detailed, quantitative description of the IAS complex as well as the
isospin-mixing.

The second example to the right in figure \ref{fig:interference} is concerned with the 6.45 MeV level and the 5.95 MeV level.
Constructive interference between these two levels can naturally explain the asymmetric shape of the 6.45 MeV level, leading to the conclusion that their spins are the same; the 5.95 MeV peak is positioned on top of a proton peak going to the $2^+$ state and its shape can not be accurately determined.
Note also that the upper edge of the 6.45 MeV level is decreasing rapidly (more than an order of magnitude before 7 MeV) which must be due to destructive interference in this region with decays through the 7.49 MeV level, an effect also included in the figure.
Interference should be visible in several other places in our spectra, such as for the two $3/2^+$ levels at 8.31 and 8.42 MeV.
The non-observation of interference effects is also significant, so the tentative 5.15 MeV level is most likely of a different spin than the 5.02 MeV level.

\section{Discussion}
\label{sec:disc}
We observe that 41(2)\% of the proton decays go to excited states in $^{20}$Ne.
Out of this roughly 2\% and 4\% of the protons go to the $4^+$ and $2^-$ states, respectively, with the rest proceeding directly to the $2^+$ state, this is in reasonable agreement with the intensity ratios derived from the gamma spectrum.
A reason for the high feeding to the $2^+$ state must be the penetrabilities that as discussed above in connection with figure \ref{fig:penet} favour transitions to it.
It is well established that $^{20}$Ne is deformed and that the mentioned states belong to the two lowest rotational bands in the nucleus.

We reinterpret peak p$_7$ in \cite{Lun15} as proceeding to the ground state in $^{20}$Ne.
The derived excitation energy is then 3859(10) keV in perfect agreement with the position of the $5/2^-$ level at 3862.2(5) keV.
Its observed intensity is 0.66(3)\% of all proton decays.
This first-forbidden branch has earlier been observed in the mirror decay of $^{21}$F \cite{Warb81}.

We were not able to measure the feeding to the $^{21}$Na ground state and therefore cannot put the branching ratios on an experimentally founded absolute scale.
To show the sensitivity to the ground state branching two evaluations are made, one where its $\log(ft)$ value is equal to that of the mirror decay of $^{21}$F, 5.67, and one where the branching is an order of magnitude lower, 6.67. This, in turn, corresponds to branching ratios of 6.1 \% and 0.6 \% to which the remaining levels' branching ratios and $\log(ft)$$_+$ values are adjusted.
Table \ref{tab:delta} gives the resulting branching ratios, the deduced $\log(ft)$ values along with those of the mirror decay (here taken from \cite{Fire15}) and the resulting difference $\Delta = $ $\log(ft)$$_+ -$ $\log(ft)$$_-$, as used in the recent overview \cite{Riis23}.
For the two first unbound levels ($5/2^+$ and $5/2^-$) the uncertainty in the branching ratio from table \ref{tab:peaks} is larger than the uncertainty induced from the ground state branching.
The overall uncertainty of the normalization procedure is of order 5--10 \%, within that most transitions agree with the mirror transitions, except for the two weakest ones.
Note that the transitions to the bound states could not be included in earlier comparisons \cite{Riis23}.

\begin{table*}
  \caption{
  Branching ratios, $b$, $\log(ft)$ values and final state excitation energy for the lowest $^{21}$Mg transitions compared to the mirror $^{21}$F transitions.
  $\Delta = $ $\log(ft)$$_+ -$ $\log(ft)$$_-$ is also listed for the different levels.
  Two different values for the ground state branch are used to display the sensitivity to this unknown quantity, see the text for further details.
  }
  \label{tab:delta}
  \begin{tabular}{lcccccc}
    \hline\noalign{\smallskip}
      level & b (\%)  & $\log(ft)$$_+$ & $\log(ft)$$_-$ & E($^{21}$Na) (keV) & E($^{21}$Ne) (keV) & $\Delta$  \\
    \noalign{\smallskip}\hline\noalign{\smallskip}
      $3/2^+$ &  6.1/0.6 & 5.67/6.67 & 5.67(16) & gs & gs & 0/$-1$ \\
      $5/2^+$ & 49.5/52.5 & 4.71/4.68 & 4.65(1) & 332 & 351 & 0.06/0.03 \\
      $7/2^+$ & 22.7/24.0 & 4.78/4.76 & 4.72(3) & 1716 & 1746 & 0.06/0.04 \\
      proton & 21.7/22.9 & n.a. & & & & \\
      $5/2^+$ & 0.33/0.34 & 6.21(6) & 7.11(5) & 3544 & 3736 & $-0.90$\\
      $5/2^-$ & 0.14/0.15 & 6.49(3) & 6.85(4) & 3862 & 3885 & $-0.36$ \\
      $5/2^+$ & 3.57/3.78 & 4.99/4.97 & 5.02(3) & 4294 & 4526 & $-0.03$/$-0.05$ \\
      $3/2^+$ & 5.03/5.32 & 4.80/4.77 & 4.5(3) & 4468 & 4685 & 0.3/0.3\\
    \noalign{\smallskip}\hline
  \end{tabular}
\end{table*}

\begin{figure}
  \includegraphics[width=0.48\textwidth]{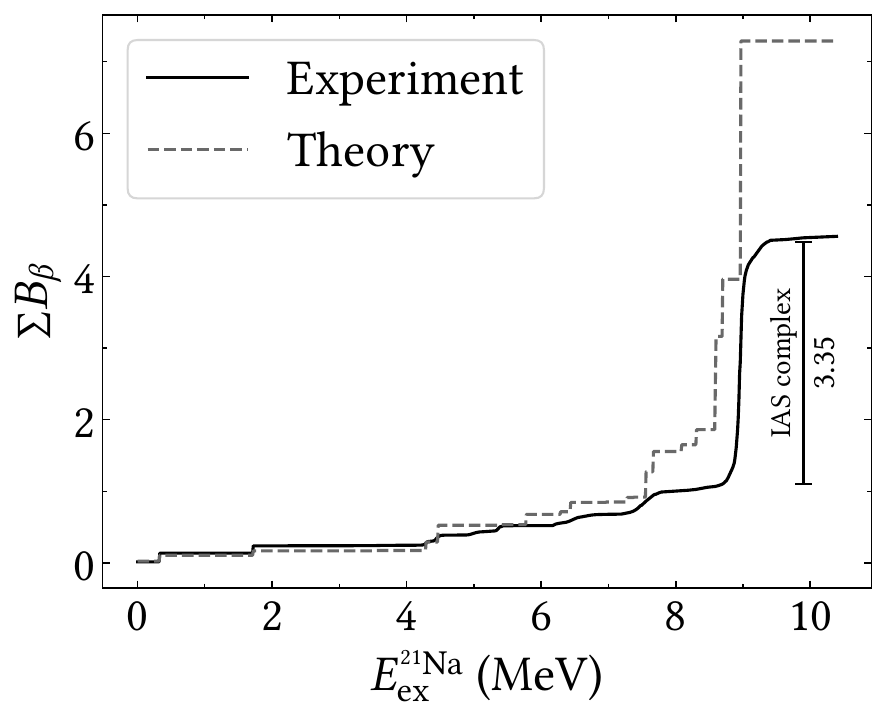}
  \caption{Deduced cumulated beta strength $\sum B_{\beta}$ of the current experiment as a function of excitation energy in $^{21}$Na, $E_{\mathrm{ex}}^{^{21}\mathrm{Na}}$.
  The experimental result is compared to theory:
  A Fermi strength of 3 units is added to the theoretical Gamow-Teller strength from \cite{Brow85} at the position of the IAS (8.976 MeV).
  The theoretical Gamow-Teller calculation is based on isospin-conserving one- and two-body interactions in an sd-space shell model.}
  \label{fig:bgt}
\end{figure}

The beta strength $\mathrm{B}_{\beta} = \mathrm{B}_F+(g_A/g_V)^2\mathrm{B}_{GT}$ can now be extracted from $\mathrm{B}_{\beta} = 6144\,\mathrm{s}/ ft$.
Since the f-factor can vary noticably across the broadest levels, it is safer to perform the conversion bin by bin in the proton spectrum (see \cite{Rii14} for further arguments why this procedure is appropriate).
Adding the intensity from the bound states and the tiny contribution from the $\beta\alpha$ decays gives the total cumulative beta strength distribution shown in fig.\ \ref{fig:bgt}.
This is compared to the theoretical calculations of the Gamow-Teller strength from \cite{Brow85} to which a Fermi strength of 3 units is added at the position of the IAS\footnote{The theoretical calculations in \cite{Brow85} are based on complete ($0d_{5/2},1s_{1/2},0d_{3/2}$)-space shell model calculations utilising shell model wave functions from \cite{Wil84} with an isospin-conserving Hamiltonian containing one- and two-body interactions.}.
The theoretical Gamow-Teller strength to the IAS is 0.33 and the total experimental strength of the IAS complex from 8.8 MeV to 9.3 MeV (adding all decay channels) is 3.35 units.
This agreement is quite good, so there are no indications for significant spread of the Fermi strength beyond the IAS complex.
The experimental and theoretical strength distributions agree well up to around 7 MeV, above this more strength is predicted than observed.
This could be due to a small displacement in the position of the Gamow-Teller Giant Resonance in the calculations or to wrongly assigned strength (in particular strength attributed to ground state rather than excited state transitions) in the experiment.

\section{Conclusion}
The experiment has improved significantly on our knowledge of the decay scheme of $^{21}$Mg.
Beta-delayed proton emission to excited states in $^{20}$Ne was confirmed to be an important decay route through explicit gamma-proton coincidences, but the fact that these decays amount to around 40\% of the total particle emission masks proton transitions to the ground state in several energy regions.
A further characteristic of the decay is that several levels in $^{21}$Na with width more than 100 keV are populated.
This leads in several cases to clear interference effects that helps in assigning spin-parity to the levels.
(Some earlier experiments wrongly interpreted the broad levels in terms of several narrow proton lines.)
A detailed fitting of these effects was hampered by the low statistics in the gamma-gated spectra.
The earlier observed $\beta$p$\alpha$ branch was confirmed and a precise half-life value was extracted.

The ground state transition could not be extracted due to the presence of directly produced $^{21}$Na in our beam.
Systematics of the ground state-feeding in the mirror decay of $^{21}$F combined with observations from the current experiment does, however, yield a branching ratio of beta-delayed proton emission in the region of 22-23\% (see table \ref{tab:delta}), which is about 10\% less than the previously stated 32.6\% in \cite{Sext73}.
To further progress on the decay scheme, it would be very valuable to get experimental values for the ground state transition.
It would e.g.\ allow a more precise comparison to the mirror decay of $^{21}$F to be made.

Much of the interpretation of the decay could be based on existing reaction experiment studies, in particular of $^{20}$Ne+p scattering.
However, there is a lack of information on (p,p') reactions that could be useful for exploring our observation that proton emission to the $2^+$ state in $^{20}$Ne in many cases are favoured to emission to the ground state.
We note finally that it would be interesting to test our tentative assignment of a new 10.70 MeV level as having T=3/2, as well as our tentative assignments of several new $7/2^+$ levels.

\begin{acknowledgements}
  We acknowledge the support of the IDS Collaboration and of the ISOLDE Collaboration and technical teams. This work has been supported by the European Commission within the Seventh Framework Programme ``European Nuclear Science and Applications Research'', contract no. 262010 (ENSAR). The authors also acknowledge the support from the Independent Research Fund Denmark, project numbers 9040-00076B and 2032-00066B, from MCIN/AEI/ 10.13039/501100011033 under grants PID2019-104390GB-I00, PID2021-126998OB-100, TED2021-130592B-100 and PID2022-140162NB-I00, from the Romanian IFA grant CERN/ISOLDE and Nucleu project No. PN 23 21 01 02, and from the United Kingdom Science and Technology Facilities Council through the grant numbers ST/P004598/1 and ST/V001027/1.
\end{acknowledgements}

\end{document}